# Edge-induced radiation force and torque on a cylindrically-radiating active acoustic source near a rigid corner-space

F.G. Mitri*

(*F.G.Mitri@ieee.org)

**Abstract**
This work examines the physical effect of the edge-induced acoustic radiation force and torque on an acoustically radiating circular source, located near a rigid corner. Assuming harmonic (linear) radiating waves of the source, vibrating in monopole or dipole radiation modes near a rigid corner-space in a non-viscous fluid, the modal series expansion method in cylindrical coordinates, the classical method of images and the translational addition theorem are applied to obtain the mathematical expressions for the radiation force and torque components in exact closed-form partial-wave series. Computational results illustrate the theory, and examine some of the conditions where the radiation force and torque components vanish, which has the potential to achieve total motion suppression (i.e., translation or rotation). Furthermore, depending on the size parameter of the source and the distances from the rigid corner space, these physical observables take positive or negative values, anticipating the prediction of pulling/pushing motions toward the corner space, and possible spinning of the source clockwise or counter-clockwise. The present analysis and its results are useful in some applications related to the manipulation of an active carrier or ultrasound contrast agents near a corner space or chamber walls at a right angle.

## I. INTRODUCTION

The presence of a corner space and edges (or boundaries) near an oscillating acoustic source affect the radiation and scattering of waves [1-6] and induce some remarkable effects of active invisibly cloaking [7, 8] and noise control. In this process, the waves emitted from the active source interact manifold with those reflected from the boundaries, and cause a physical phenomenon of zero extinction, in addition to either positive or negative extinction depending on the size of the source and the distances from the boundaries.

During this process, adequate analytical modeling is needed in order to fully understand the physical nature of the encountered phenomena and the necessary conditions so as to achieve such effects. An important question that arises is to which extent the presence of the corner space (see Fig. 1 in [8]) will affect the acoustic radiation force and torque on the source due to multiple reflections effects with the boundaries. The aim of this work is to address this challenge from the standpoint of acoustic radiation force and torque theories, stemming from an analysis of the radiated field examined recently using the multipole expansion in partial wave series expansions [8]. It is anticipated that the present exact analytical formalism can be used to predict some of the conditions for the edge-induced radiation force and torque on an active carrier (i.e. acoustic source), for example, in ultrasound contrast agent applications, and other related areas of research related to the manipulation of active cylindrical/elongated particles in fluid dynamics.

For the purpose of the present study, the expressions for the radiated fields from the circular cylindrical source nearby a rigid corner (i.e., quarter-space) and its images developed in [8] are directly used and noted in the text, without loss of generality. After determining the appropriate expansion coefficients for the source and its three images, the expressions for the radiated velocity potential fields are used to derive the edge-induced radiation force and torque components for the active source in cylindrical coordinates. Then, numerical simulations are considered with particular emphases on the source size and the distances from the rigid corner space.

## II. DERIVATION OF THE EDGE-INDUCED RADIATION FORCE AND TORQUE

Stemming from the general mathematical equation [9], it is applied here to derive the radiation force vector (per length) experienced by the cylindrical source as,

$$\langle \mathbf{F} \rangle_{kr_1 \to \infty} = \frac{1}{2} \rho k^2 \int_0^{2\pi} \Re\{\Phi_{is}\} d\mathbf{S}, \quad (1)$$

where $\Re\{\cdot\}$ is the real part of a complex number, $k$ is the wave number in the host medium, $\rho$ is the mass density of the fluid medium surrounding the object, and the differential surface vector of a cylindrical surface enclosing the active source is $d\mathbf{S} = dS\,\mathbf{n}$, where $dS = r_1 d\theta_1$. The normal unit vector is expressed as $\mathbf{n} = \cos\theta_1 \mathbf{e}_x + \sin\theta_1 \mathbf{e}_y$, where $\mathbf{e}_x$ and $\mathbf{e}_y$ are the unit vectors, and the symbol $\langle \cdot \rangle$ denotes time-averaging.

Based on the analysis established in [8], the factor $\Phi_{is}$ appearing in Eq.(1) is expressed as

$$\Phi_{is} = \Phi_{rad}^{1*}\left[(i/k)\partial_r\left(\sum_{\alpha=2}^{4}\Phi_{rad}^{\alpha}\right) - \sum_{\alpha=2}^{4}\Phi_{rad}^{\alpha} - \Phi_{rad}^{1}\right] \quad (2)$$

where the expressions for $\Phi_{rad}^1$ and $\Phi_{rad}^{\alpha\neq 1}$ are given by Eqs.(1) and (4) in [8], respectively, and the superscript * indicates the conjugate of a complex number.

The substitution of Eq. (2) into Eq.(1) using the far-field expressions (i.e., $kr_1 \to \infty$) of Eqs.(1) and (4) in [8], leads to the determination of two radiation force components along the axes $\mathbf{x_1}$ and $\mathbf{y_1}$ (see Fig. 1 in [8]), as,

$$\begin{Bmatrix} F_x \\ F_y \end{Bmatrix} = \langle \mathbf{F} \rangle \cdot \begin{Bmatrix} \mathbf{e}_x \\ \mathbf{e}_y \end{Bmatrix} = \begin{Bmatrix} Y_x \\ Y_y \end{Bmatrix} S_c E_0. \quad (3)$$

In Eq.(3), $E_0 = \tfrac{1}{2}\rho k^2 |\phi_0|^2$ is a characteristic energy density factor, the parameter $S_c = 2a$ is the surface cross-section per unit length of the active source, and $Y_x$ and $Y_y$ are the non-dimensional longitudinal and transverse radiation force functions. Their expressions are obtained after some algebraic manipulation, respectively, as

$$\boxed{Y_x = \frac{1}{ka}\Im\left\{\sum_{n=-\infty}^{+\infty}\left(C_n^1 + \sum_{\alpha=2}^{4}D_{nm}^{\alpha 1}\right)\left(C_{n+1}^{1*} - C_{n-1}^{1*}\right)\right\}}, \quad (4)$$

$$\boxed{Y_y = -\frac{1}{ka}\Re\left\{\sum_{n=-\infty}^{+\infty}\left(C_n^1 + \sum_{\alpha=2}^{4}D_{nm}^{\alpha 1}\right)\left(C_{n+1}^{1*} + C_{n-1}^{1*}\right)\right\}}, \quad (5)$$

where $\Im\{\cdot\}$ denotes the imaginary part of a complex number, and the coefficients $D_{nm}^{\alpha 1}$ appearing in Eqs.(4) and (5) are given explicitly by Eq.(5) of [8]. Those coefficients depend on the distances separating the source from its images as well as the corresponding angles (see Fig. 1 in [8]).

The analysis is extended to derive the expression for the edge-induced radiation torque exerted on the active cylindrical source of circular cross-section, located near the rigid corner space. In 2D, the non-vanishing axial component (along the z-direction) is obtained stemming from the standard expression [10] (applied also for a cylindrically-focused nonparaxial Gaussian beam[11], a pair of fluid cylinders [12] and plane waves on an elliptical cylinder [13]) given by,

$$\langle \mathbf{N} \rangle_{kr_1 \to \infty} = -\rho \int_0^{2\pi} \langle \mathbf{v} \otimes (\mathbf{r_1} \times \mathbf{v}) \rangle \cdot d\mathbf{S}, \quad (6)$$

where the symbol $\otimes$ denotes a tensor product. In Eq.(6), the total velocity vector in the system of coordinates $(r_1,\theta_1)$ is expressed as, $\mathbf{v} = \nabla\left(\Phi_{rad}^1(r_1,\theta_1,t) + \sum_{\alpha=2}^{4}\Phi_{rad}^{\alpha}(r_1,\theta_1,t)\right)$,

where the expressions for $\Phi_{rad}^1$ and $\Phi_{rad}^{\alpha\neq 1}$ are given by Eqs.(1) and (4) in [8], respectively. Moreover, the only non-vanishing (axial) component of the time-averaged radiation torque vector [i.e. in the direction perpendicular to the polar plane $(r_1,\theta_1)$] can be rewritten as,

$$N_z^{rad} = \langle \mathbf{N} \rangle \cdot \mathbf{e}_z$$
$$= \frac{\rho}{2}\Im\left\{\int_0^{2\pi} \frac{\partial}{\partial r_1}\left(\Phi_{rad}^1(r_1,\theta_1,t) + \sum_{\alpha=2}^{4}\Phi_{rad}^{\alpha}(r_1,\theta_1,t)\right)^* \times \hat{L}_z\left(\Phi_{rad}^1(r_1,\theta_1,t) + \sum_{\alpha=2}^{4}\Phi_{rad}^{\alpha}(r_1,\theta_1,t)\right)dS\right\}, \quad (7)$$

where $\mathbf{e}_z$ is the unit vector along the z-direction, and $\hat{L}_z$ is the z-component of the angular momentum operator in polar coordinates given by,

$$\hat{L}_z = -i\frac{\partial}{\partial \theta_1}. \quad (8)$$

Using the expressions for $\Phi_{rad}^1$ and $\Phi_{rad}^{\alpha\neq 1}$ in the system of coordinates $(r_1,\theta_1)$, which were given previously by Eqs.(1) and (4) in [8], Eq.(7) leads to the axial radiation torque component as

$$N_z^{rad} = \tau_z V E_0, \quad (9)$$

where $V = \pi a^2$ is the volume of the cylindrical source of unit-length, and $\tau_z$ is the non-dimensional axial radiation torque function, obtained as

$$\boxed{\tau_z = -\frac{4}{\pi(ka)^2}\Re\left\{\sum_{n=-\infty}^{+\infty} n C_n^{1*}\left(C_n^1 + \sum_{\alpha=2}^{4}D_{nm}^{\alpha 1}\right)\right\}}. \quad (10)$$

Eqs. (3)-(5), (9) and (10) constitute the main contribution of this work, which are suitable to predict the edge-induced acoustic radiation force and torque vectors from any radiating source of arbitrary shape in 2D located near a rigid corner space. In the following, computational examples are considered for an oscillating cylindrical source of circular cross-section in order to illustrate the theory and examine some of the conditions where singularities can arise (i.e., zero force and torque) leading to a complete unresponsiveness of the source to the transfer of linear and angular momenta. Numerical implementations of Eqs.(4), (5) and (10) are performed, after determining the expansion coefficients $C_n^1$ and $D_{nm}^{\alpha 1}$ appearing in those equations, following the method presented recently in [8]. Emphases are given on varying the oscillatory mode order of the acoustic source, its size parameter and the dimensionless distances from the corner.

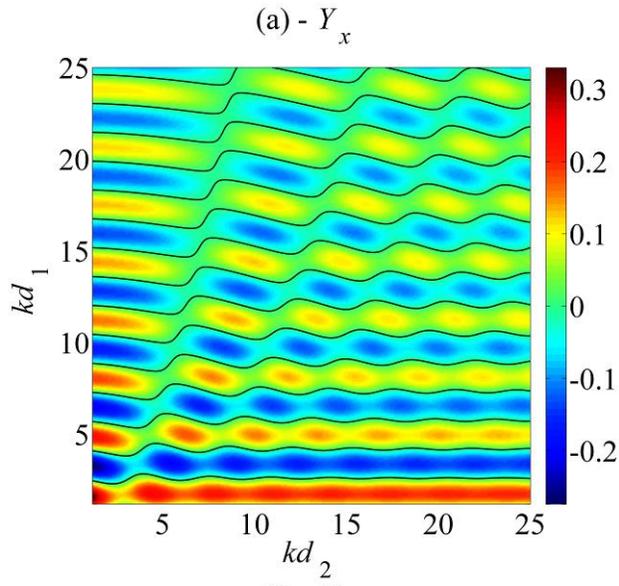

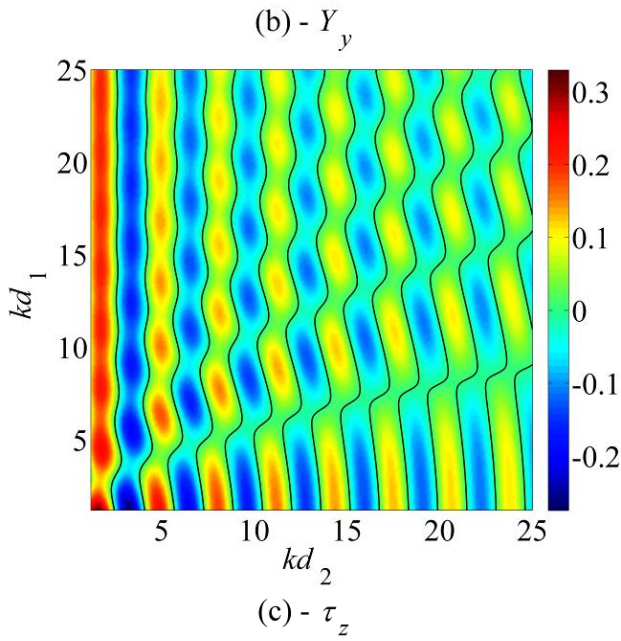

**Fig. 1.** Panels (a)-(c) display the plots for the longitudinal and transversal radiation force functions, in addition to the axial radiation torque function, respectively, assuming a monopole ($n = 0$) radiation mode for a Rayleigh source having $ka = 0.1$.

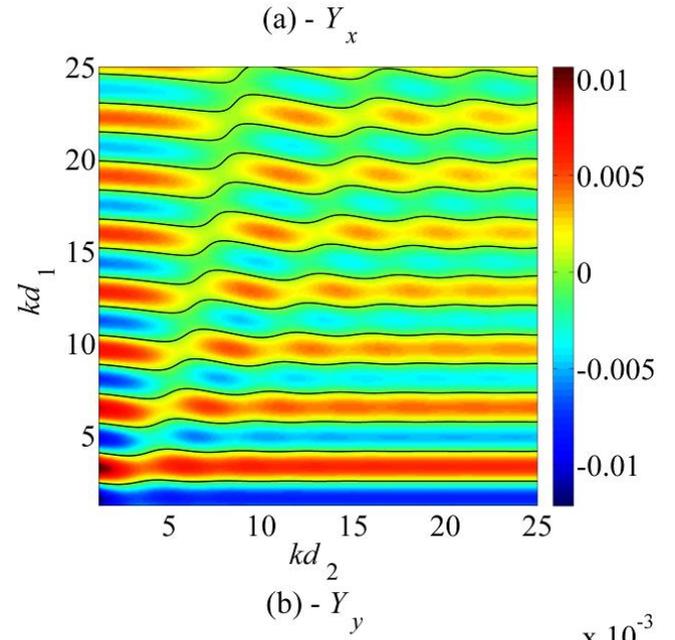

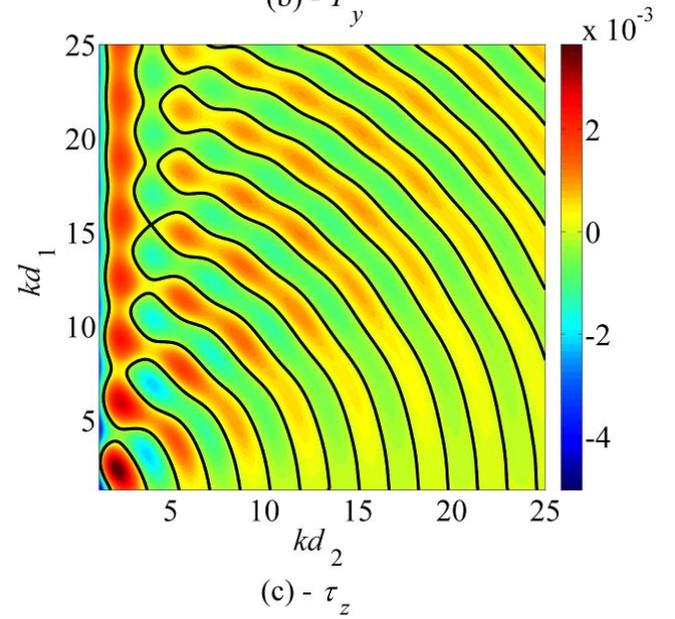

**Fig. 2.** The same as in Fig. 1, but for a dipole ($|n| = 1$) radiating source.

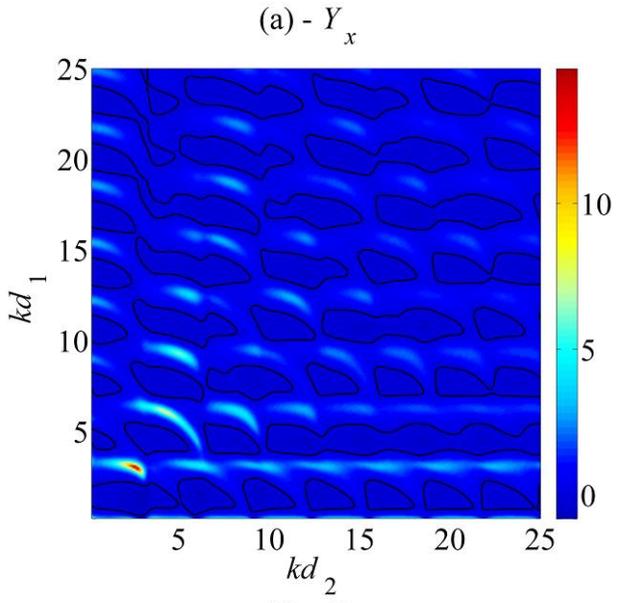

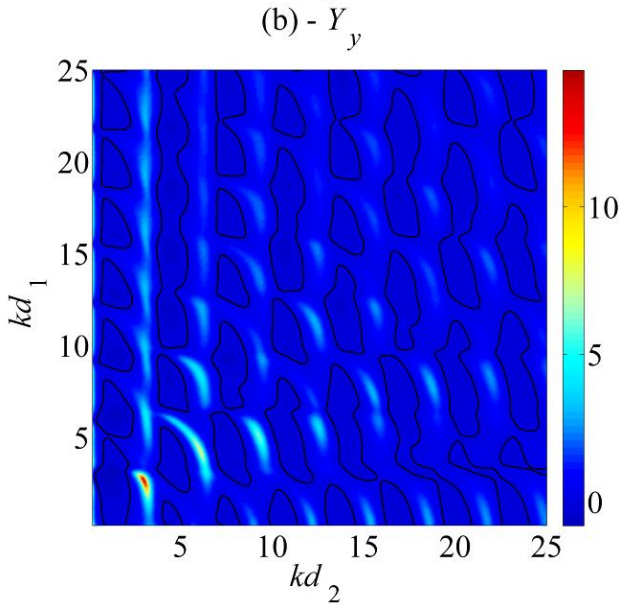

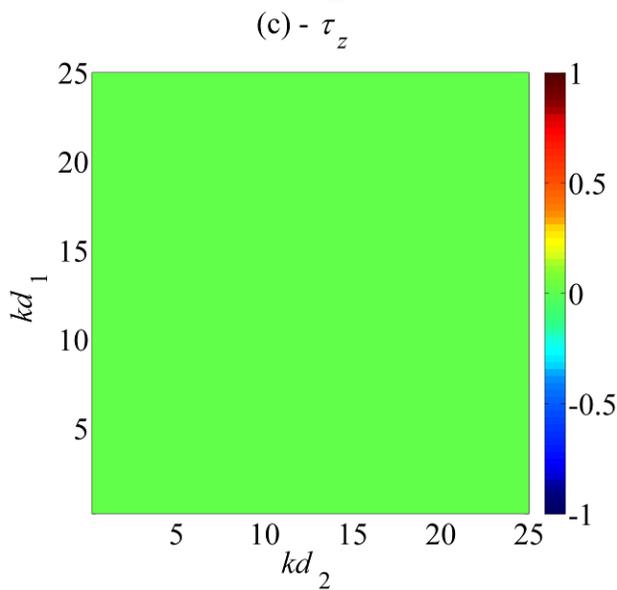

**Fig. 3.** The same as in Fig. 1, but $ka = 5$.

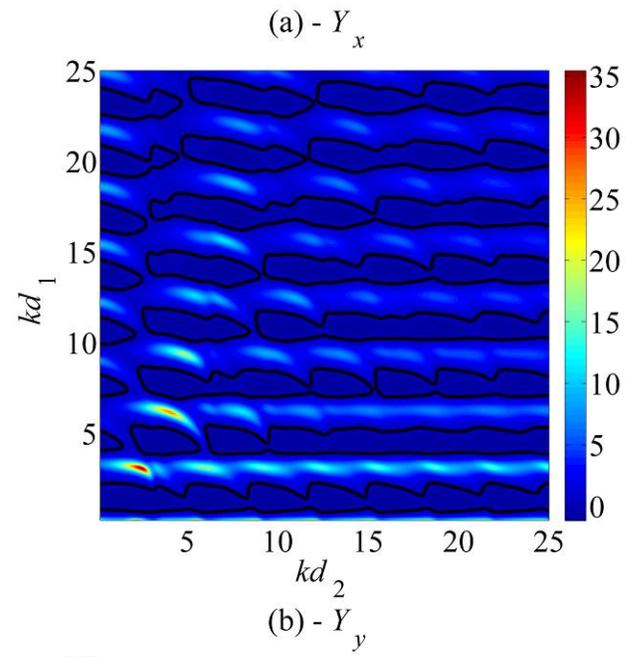

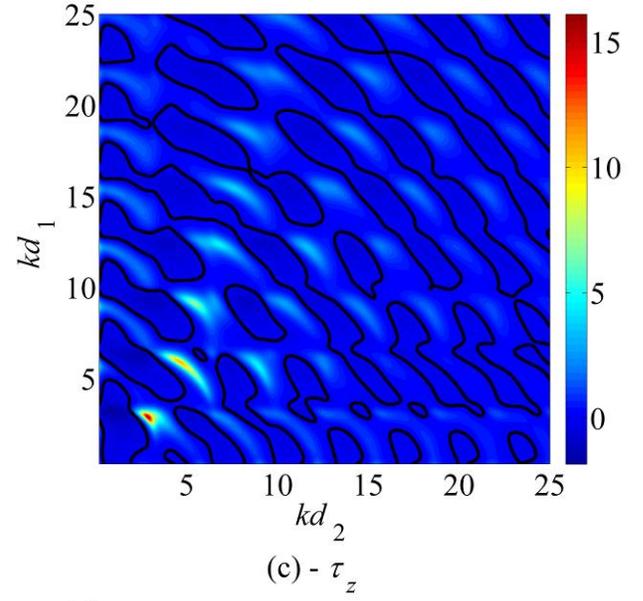

**Fig. 4.** The same as in Fig. 2, but $ka = 5$.

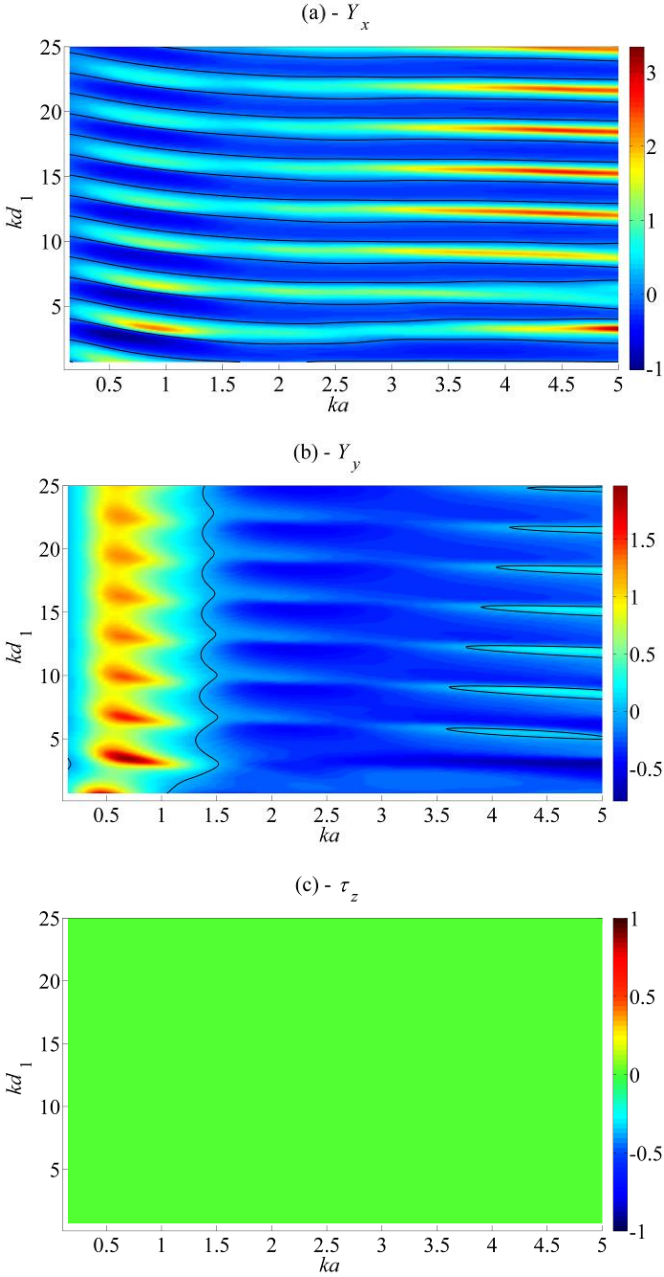

**Fig. 5.** Panels (a)-(c) display the plots for the longitudinal and transversal radiation force functions, in addition to the axial radiation torque function, respectively, assuming a monopole ($n = 0$) radiation mode at $kd_2 = 1$.

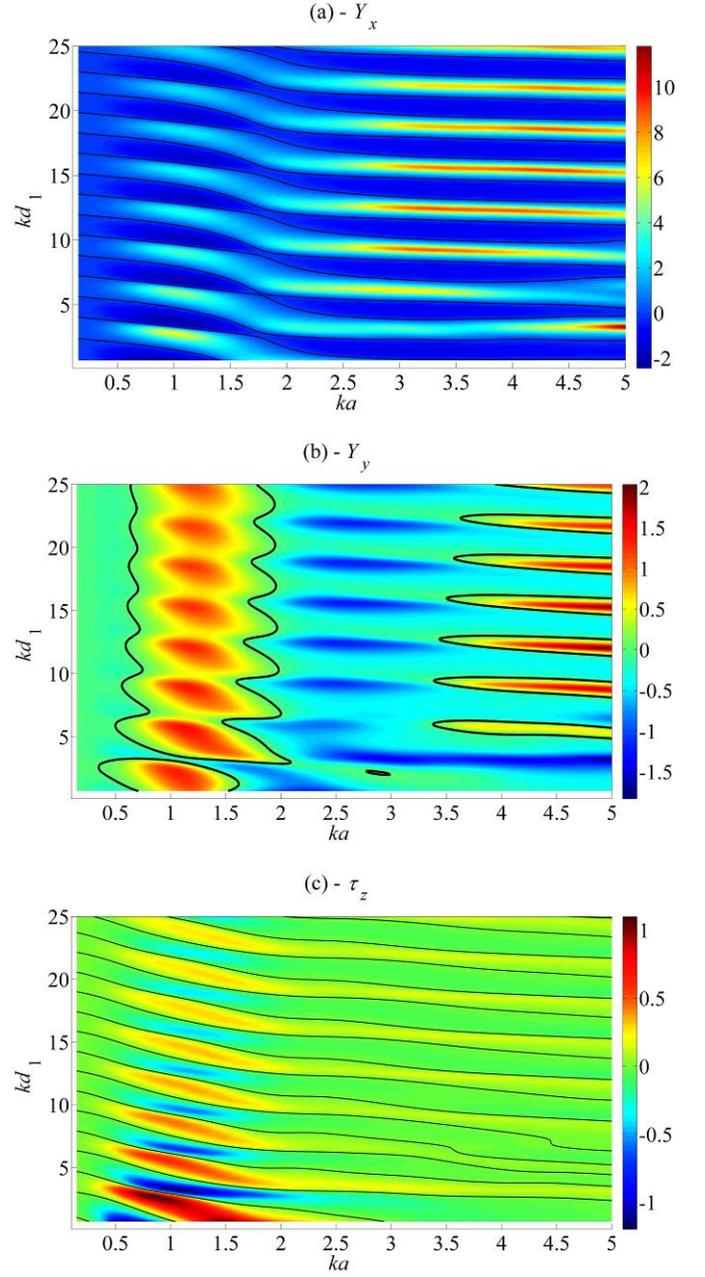

**Fig. 6.** The same as in Fig. 5, but for a dipole ($|n| = 1$) oscillating source.

## III. NUMERICAL EXAMPLES AND DISCUSSIONS

Panels (a)-(c) of Fig. 1 display the computational results for the non-dimensional longitudinal $Y_x$ and transversal $Y_y$ radiation force components, as well as the axial radiation function $\tau_z$, respectively, in the ranges $0 < (kd_1, kd_2) \leq 25$ for a Rayleigh (i.e., small) cylindrical source with $ka = 0.1$, assuming a monopole ($n = 0$) oscillatory mode. The monopole vibration corresponds to a breathing mode with radial symmetry with respect to the corner-space. As shown in panel (a), the components $Y_x$ varies between either positive or negative values, such that the active radiating source can be attracted toward the rigid quarter-space, or gets repelled away from it depending on the values of $kd_1$ and $kd_2$. The solid (black) contour indicates the lines of singularity over which $Y_x = 0$, and the source does not experience any edge-induced longitudinal force component. Correspondingly, panel (b) displays the result for $Y_y$, where a similar behavior occurs. Along the lines of singularity, the active source becomes irresponsive to the transfer of linear momentum caused by the waves interacting multiple times between the source and quarter-space. Moreover, due to the present of the corner, there exists a correlation between the radiation force components for this type of monopole oscillatory vibration, such that $Y_x(kd_1, kd_2) = Y_y(kd_2, kd_1)$. Should the lines of singularity for the longitudinal and transversal radiation force functions overlap, a complete immobilization of the active source can occur near the rigid corner-space. These results can help explaining why an oscillating particle (for example, an active ultrasound contrast agent or bubble) near a corner space becomes trapped therein. Furthermore, panel (c) shows that the axial component of the radiation torque function vanishes due to the nature of the monopole mode presenting symmetry with respect to the corner.

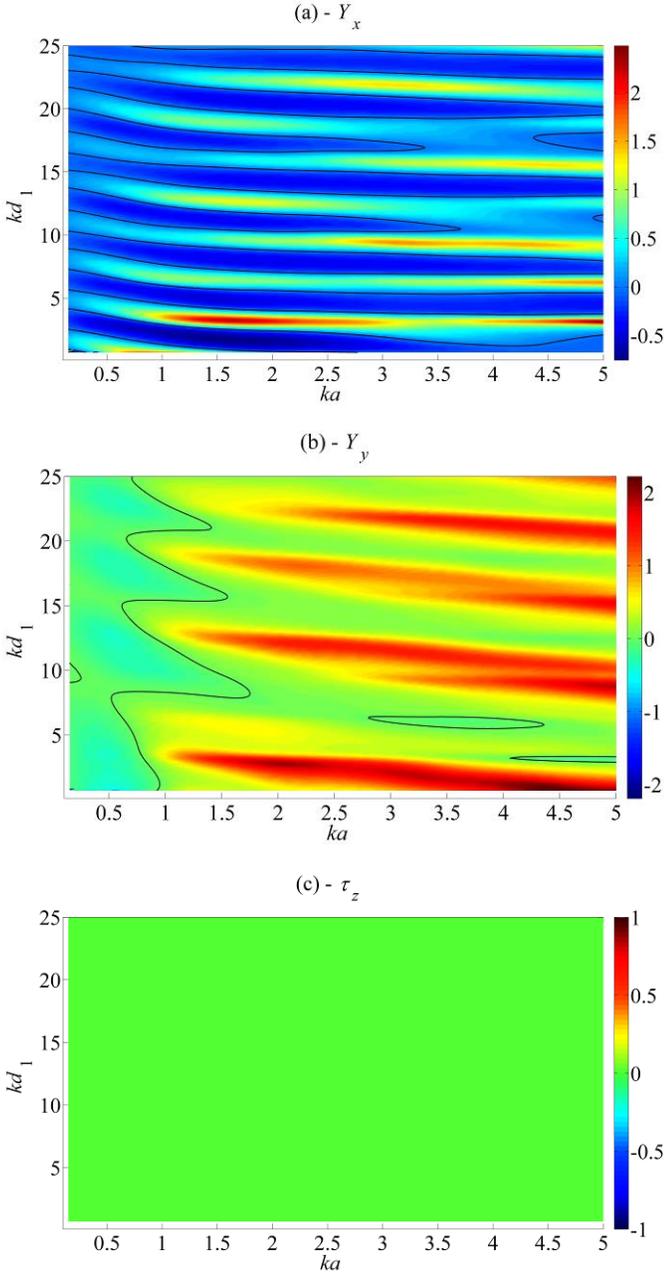

**Fig. 7.** The same as in Fig. 5, but $kd_2 = 25$.

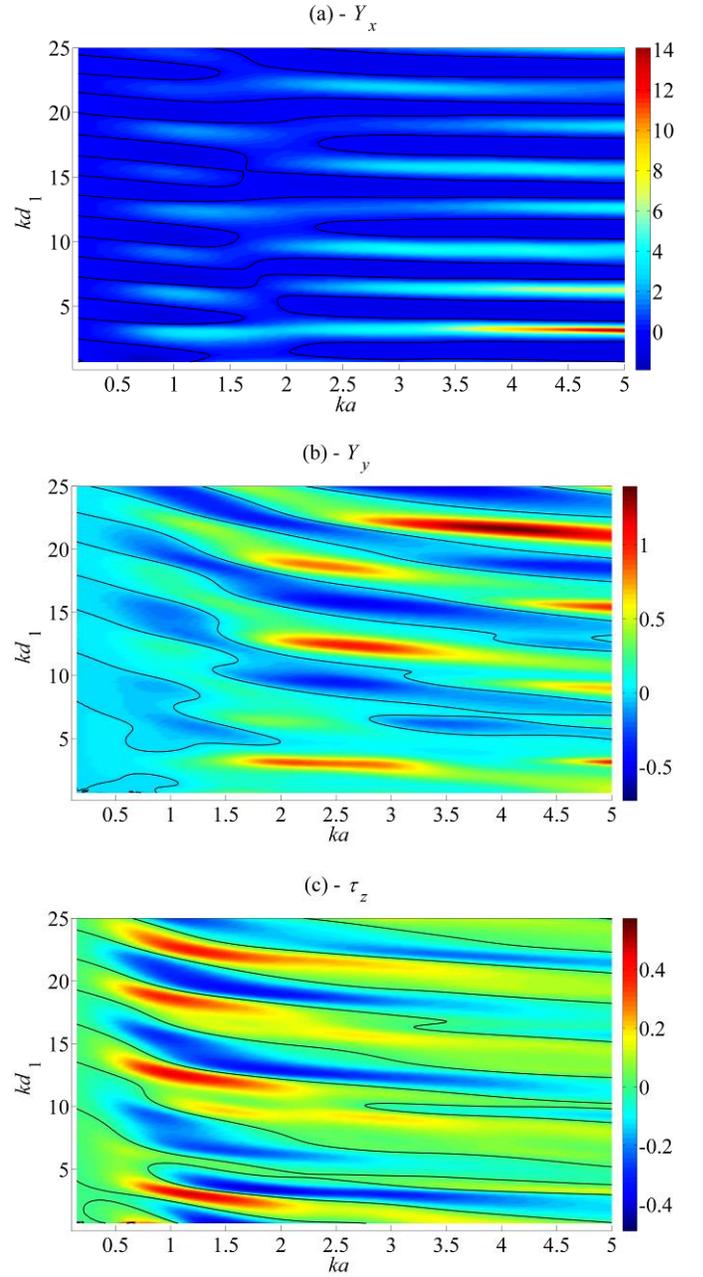

**Fig. 8.** The same as in Fig. 6, but $kd_2 = 25$.

The effect of changing the vibrational mode order is further analyzed by considering now an asymmetric dipole ($|n| = 1$) back-and-forth oscillation of the Rayleigh source with $ka = 0.1$. The results are shown in panels (a)-(c) of Fig. 2, where the relationship between $Y_x$ and $Y_y$ observed previously for the monopole mode in panels (a) and (b) of Fig. 1 is now lacking. As shown in Fig. 1, panels (a) and (b) of Fig. 2 demonstrate that $Y_x$ and $Y_y$ alternate between positive or negative values, and can vanish along well-defined lines of singularity depending on the choice of $kd_1$ and $kd_2$. Panel (c) also shows that $\tau_z$ alternates between positive or negative values depending on $kd_1$ and $kd_2$, leading to a possible rotation of the dipole oscillating source in the counter-clockwise or the clockwise, respectively. Also, axial torque lines of singularity arise where the source becomes non-responsive to the angular momentum transfer from the waves reflected manifold from the corner. One notices the symmetry of the plot in panel (c) with respect to the diagonal line $kd_1 = kd_2$.

Increasing the size parameter of the source is now investigated such that $ka = 5$, and panels (a)-(c) of Fig. 3 display results for a monopole vibrating source, which are significantly different from those computed for the Rayleigh source shown in Fig. 1. As $ka$ increases, $Y_x$ in panel (a) continues to alternate between positive and negative values, suggesting the effect of the source pushing toward the corner space or its repulsion away from it. Nonetheless, the singularity lines become discontinuous and form well-defined "islands" of negative $Y_x$. A similar behavior is also observed in the plot for $Y_y$ in panel (b), where the relationship for the monopole vibration mode still holds; that is, $Y_x(kd_1, kd_2) = Y_y(kd_2, kd_1)$. Notice also that the amplitudes of the radiation force functions $Y_x$ or $Y_y$ are larger (in the absolute sense) as $ka$ increases. As for the axial torque component, $\tau_z$ vanishes as shown in panel (c), $\forall$ the value of $ka$.

Panels (a)-(c) of Fig. 4 correspond to the results assuming a dipole vibration of the source with $ka = 5$. Visual comparison of each of the panels with those of Fig. 2 for the Rayleigh source shows that the increase in $ka$ significantly affects the emergence of positive or negative force and torque components and alters the lines of singularity. The arc pattern shown in panel (c) of Fig. 2 becomes of a more complex form as shown in panel (c) of Fig. 4, maintaining the symmetry in the plot with respect to the diagonal line $kd_1 = kd_2$.

Consider now the variations of the radiation force and torque components versus the size parameter and dimensionless distance in the ranges $0 < ka \leq 5$ and $0 < kd_1 \leq 25$, for fixed values of $kd_2$. Panels (a)-(c) of Fig. 5 correspond to the plots of $Y_x$, $Y_y$ and $\tau_z$ assuming a monopole radiating source for $kd_2 = 1$. As $ka$ and $kd_1$ increase, $Y_x$ and $Y_y$ alternate between positive and negative values while they vanish along the lines of singularity. The plot for $Y_x$ in panel (a) displays a distinctive pattern quite different from that of panel (b), while $\tau_z$ vanishes for this kind of source vibration.

Changing the source vibration to a dipole oscillation alters the signs of $Y_x$, $Y_y$ and $\tau_z$ as demonstrated in panels (a)-(c) of Fig. 6. Moreover, with this type of back-and-forth oscillatory motion of the source, $\tau_z$ alternates between positive and negative values and vanishes along the solid lines of singularity.

As the non-dimensional distance $kd_2$ increases, changes occur in the plots of $Y_x$, $Y_y$ and $\tau_z$ as shown in the panels of Figs. 7 and 8, respectively, for $kd_2 = 25$.

## IV. CONCLUSION

In summary, this works demonstrates that an active cylindrical source in a non-viscous fluid vibrating near a rigid corner experiences a acoustic radiation force and torque, which can cause its translation and rotation. Exact closed-form series expansions are derived and computed numerically with particular emphases on the source size, its vibrational mode shape and the distances from the rigid corner space. The results show that attractive or repulsive forces can arise as well as possible counter-clockwise or clockwise rotations of the active source are anticipated. Assuming a monopole radiation of the source, no torque can be generated due to symmetry considerations. Numerical computations illustrate the analysis where some of the physical properties for active radiation force and torque cloaking are examined. In particular, the results demonstrate the emergence of singularity lines where total non-responsiveness to the linear and angular momenta transfer can arise. Such an effect may be used to advantage in the particle manipulation of an acoustically-active carrier or contract agent. Although the results are presented here for a cylindrical source in 2D, it is anticipated that similar behaviors can emerge for the 3D spherical particle, and this work should help in developing other models for other source geometries.